# Hierarchical Fuzzy Opinion Networks: Top-Down for Social Organizations and Bottom-Up for Election

Li-Xin Wang

*Abstract*—A fuzzy opinion is a Gaussian fuzzy set with the center representing the opinion and the standard deviation representing the uncertainty about the opinion, and a fuzzy opinion network is a connection of a number of fuzzy opinions in a structured way. In this paper, we propose: (a) a top-down hierarchical fuzzy opinion network to model how the opinion of a top leader is penetrated into the members in social organizations, and (b) a bottom-up fuzzy opinion network to model how the opinions of a large number of agents are agglomerated layer-by-layer into a consensus or a few opinions in the social processes such as an election. For the top-down hierarchical fuzzy opinion network, we prove that the opinions of all the agents converge to the leader's opinion, but the uncertainties of the agents in different groups are generally converging to different values. We demonstrate that the speed of convergence is greatly improved by organizing the agents in a hierarchical structure of small groups. For the bottom-up hierarchical fuzzy opinion network, we simulate how a wide spectrum of opinions are negotiating and summarizing with each other in a layer-by-layer fashion in some typical situations.

*Index Terms*—Opinion dynamics; social hierarchy; fuzzy opinion networks.

## I. Introduction

Hierarchy is the most popular structure in social organizations such as government, army, company, etc. [1,2,3]. In a company, for example, it is typically structured with a relatively small top management team, at least one layer of middle management, and a large number of lower level employees responsible for day-to-day operations [4]. Why is hierarchy so pervasive in human societies across almost all cultures throughout time [5], giving the fact that hierarchy is in direct opposition to some of the best ideas humanity has produced such as democracy, equality, fairness, and justice [1] ? The interdisciplinary research on social hierarchy in sociology [6,7], psychology [8,9], management [10], economics [11], and other disciplines suggest a number of reasons. First, hierarchy establishes social order that is appealing psychologically to individuals who need safety and stability (people come and go, but the system remains). Second, hierarchy provides incentives for individuals in organizations to work hard to obtain higher rank to satisfy material self-interest and their need for control that in turn serves the organization's interests (motivating the individuals to work hard for the organization). Third, hierarchy facilitates coordination and improves efficiency in comparison to other more egalitarian structures such as free markets (the rapid development of China's "state capitalism" economy in recent years is an example, which has led to the deglobalization movement in the "laisser-faire capitalism" economies to reestablish the hierarchy). Fourth, hierarchical differentiation between people fosters more satisfying working relationships (leaders provide the guidance their followers need, and followers execute what the leaders want to be realized).

Although hierarchy has been studied in social sciences for a long time (back to Marx and Engels [12] in 1846), the research is largely qualitative without mathematical modeling. Usually, some concepts or variables are defined verbally, then a theory is developed that describes the relationships among these variables using natural languages rather than mathematical equations [13]. The most mathematically advanced study related to social hierarchy is perhaps the multidisciplinary field of opinion dynamics where the researchers from mathematical sociology [14], economics [15,16], physics [17,18], social psychology [19], control [20], signal processing [21], fuzzy systems [22], etc., join forces to tackle the problem. A shortcoming of the mainstream opinion dynamics models [23,24] is that the uncertainties of the opinions are not included in the models. Human opinions are inherently uncertain so that an opinion and its uncertainty should be considered simultaneously to give the accurate picture of the opinion. For example, when we are asked to review a research paper, we need to give an overall rating for the paper and, at the same time, we must claim our level of expertise on the subject which is a measure of uncertainty about the overall rating. In fact, the uncertainty may be more important than the opinion itself in many situations, because the uncertainty is more directly related to the psychological pressure of the agent when the opinion is broadcasted [25].

Li-Xin Wang is with the University of Chinese Academy of Sciences, Beijing, P.R. China (e-mail: lxwang@ucas.edu.cn).

The fuzzy opinion networks (FONs) proposed in [22] model an opinion by a Gaussian fuzzy set whose center and standard deviation represent the opinion and its uncertainty, respectively, so that the interactions between the opinions and their uncertainties are systematically exploited. The goal of this paper is to use the FON framework to model social hierarchy. According to [2], social hierarchies can be classified into two types: i) formal hierarchies that are delineated by rule and consensually agreed upon, and ii) informal hierarchies that are established and subjectively understood during the interaction among social members. Formal hierarchies are top-down – a hierarchical structure is designed first and members at different levels are then recruited; informal hierarchies are bottom-up – a hierarchical structure emerges after the free interaction among the community members. We will use fuzzy opinion networks to model both formal and informal hierarchies. To model the formal hierarchy, we first define a basic leader-follow group of fuzzy agents and study its basic convergence properties, then we connect the basic leader-follower groups in a hierarchical fashion to get the final hierarchical fuzzy opinion network. To model an informal hierarchy, we let a large number of fuzzy agents to interact with each other based on a local reference scheme, and we see the initially very diversified opinions are merging gradually in a hierarchical fashion into a consensus or a number of representative opinions – a process very similar to an election in a democratic society.

This paper is organized as follows. In Section II, the top-down hierarchical fuzzy opinion networks are constructed and their convergence properties are proved. We also show that the speed of convergence to the top leader's opinion is greatly improved by organizing the followers into a hierarchical structure rather than in a flat nonhierarchical fashion. In Section III, we construct the bottom-up hierarchical fuzzy opinion networks through the natural process of free interacting among a large number of fuzzy agents based on the local reference scheme, and we simulate a number of typical scenarios – consensus reaching, polarization, or converging to multiple ends. Finally, Section IV concludes the paper and the Appendix contains the proofs of the theorems in the paper.

## II. Top-Down Hierarchical Fuzzy Opinion Networks

We start with the definition of fuzzy opinion networks (FON) and bounded confidence FONs[1], and then introduce the basic leader-follower group which is the basic building block of the top-down hierarchical fuzzy opinion networks.

**Definition 1:** A *fuzzy opinion* is a Gaussian fuzzy set $X$ with membership function $\mu_X(x) = e^{-\frac{|x-c|^2}{\sigma^2}}$ where the center $c \in R$ represents the opinion and the standard deviation $\sigma \in R_+$ characterizes the uncertainty about the opinion $c$. A *Fuzzy Opinion Network* (FON) is a connection of a number of Gaussian nodes, where a *Gaussian node* is a 2-input-1-output fuzzy opinion $X_i$ with Gaussian membership function $\mu_{X_i}(x) = e^{-\frac{|x-C_i|^2}{\Sigma_i^2}}$ whose center $C_i$ and standard deviation $\Sigma_i$ are two input fuzzy sets to the node and the fuzzy set $X_i$ itself is the output of the node. A Gaussian node is also called an *agent*, a *node*, or a *fuzzy node* throughout this paper. ∎

The connection of the fuzzy nodes can be static or dynamically changing with time and the status of the nodes. The bounded confidence fuzzy opinion networks, defined below, are FONs with connections that are dynamically changing according to the states of the nodes – if the fuzzy opinions of two nodes are close enough to each other, they are connected; otherwise, they are disconnected.

**Definition 2:** A *bounded confidence fuzzy opinion network* (BCFON) is a dynamic connection of $n$ fuzzy nodes $X_i(t)$ ($i = 1,2,...,n$) with membership functions $\mu_{X_i(t)}(x) = e^{-\frac{|x-C_i(t)|^2}{u_i^2(t)}}$, where the center input $C_i(t+1)$ and the standard deviation input $u_i(t+1)$ to node $i$ at time $t+1$ ($t = 0,1,2,...$) are determined as follows: the center input $C_i(t+1)$ is a weighted average of the outputs $X_j(t)$ of the $n$ fuzzy nodes at the previous time point $t$:

$$C_i(t+1) = \sum_{j=1}^{n} w_{ij}(t) X_j(t) \quad (1)$$

with the weights

$$w_{ij}(t) = \begin{cases} \frac{1}{|N_i(t)|}, & j \in N_i(t) \\ 0, & j \notin N_i(t) \end{cases} \quad (2)$$

where $N_i(t)$ ($i = 1,...,n$) is the collection of nodes that are connected to node $i$ at time $t$, defined as:

$$N_i(t) = \{j \in \{1,...,n\} \mid \max(X_i(t) \cap X_j(t)) \geq d_i\} \quad (3)$$

where $\max(X_i(t) \cap X_j(t))$ represents the closeness between fuzzy opinions $X_i(t)$ and $X_j(t)$, $d_i \in [0,1]$ are constants and $|N_i(t)|$ denotes the number of elements in $N_i(t)$; and, the standard deviation input $u_i(t+1)$ are determined according to one of the two schemes:

(a) *Local reference scheme*:

$$u_i(t+1) = b \left| \bar{x}_i(t) - \frac{1}{|N_i(t)|} \sum_{j \in N_i(t)} (\bar{x}_j(t)) \right| \quad (4)$$

(b) *External reference scheme*:

$$u_i(t+1) = b|\bar{x}_i(t) - g_i(t)| \quad (5)$$

where $\bar{x}_i(t)$ denotes the center of fuzzy set $X_i(t)$, $g_i(t)$ is an external signal and $b$ is a positive scaling constant. The initial fuzzy opinions $X_i(0)$ ($i = 1,2,...,n$) are Gaussian fuzzy sets $\mu_{X_i(0)}(x) = e^{-\frac{|x-\bar{x}_{i0}|^2}{\sigma_{i0}^2}}$, where the initial opinions $\bar{x}_{i0} \in R$ and the initial uncertainties $\sigma_{i0} \in R_+$ are given. ∎

---

[1] The concept of FON was introduced in [22] and the basic convergence properties of bounded confidence FONs were studied in [26].

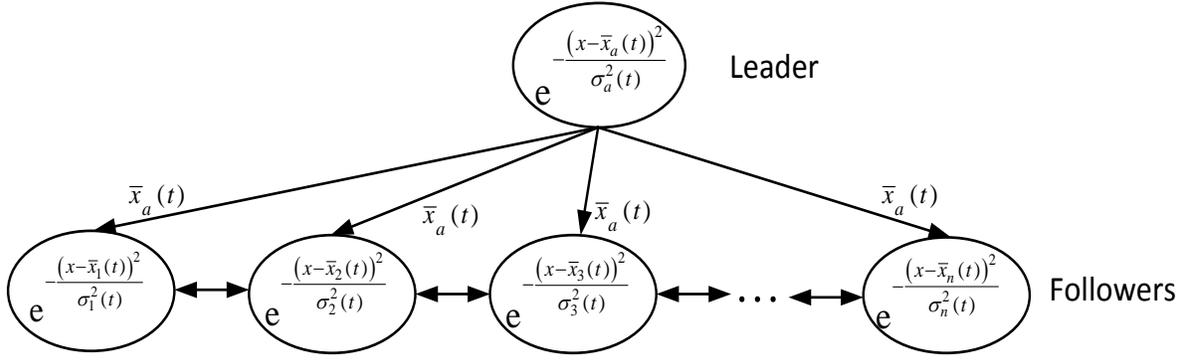

Fig. 1: The basic leader-follower group, where the leader passes his opinion $\bar{x}_a(t)$ to each of the $n$ followers who are connected among themselves in the bounded confidence fashion.

It was proved in [26] that the opinions $\bar{x}_i(t)$ and their uncertainties $\sigma_i(t)$ of the Gaussian nodes $X_i(t)$ in the BCFON are evolving according to the following dynamic equations.

**The Evolution of BCFON:** The fuzzy opinions $X_i(t)$ ($i = 1, \ldots, n; t = 0, 1, 2, \ldots$) in the BCFON are Gaussian fuzzy sets:

$$\mu_{X_i(t)}(x) = e^{-\frac{|x-\bar{x}_i(t)|^2}{(\sigma_i(t))^2}} \tag{6}$$

where the opinions $\bar{x}_i(t) \in R$ and their uncertainties $\sigma_i(t) \in R_+$ are evolving according to the following dynamic equations:

$$\bar{x}_i(t+1) = \sum_{j=1}^{n} w_{ij}(t)\bar{x}_j(t) \tag{7}$$

$$\sigma_i(t+1) = \sum_{j=1}^{n} w_{ij}(t)\sigma_j(t) + u_i(t+1) \tag{8}$$

where the weights

$$w_{ij}(t) = \begin{cases} \frac{1}{|N_i(t)|}, & j \in N_i(t) \\ 0, & j \notin N_i(t) \end{cases} \tag{9}$$

$$N_i(t) = \left\{ j \in \{1, \ldots, n\} \mid e^{-\frac{|\bar{x}_i(t)-\bar{x}_j(t)|^2}{(\sigma_i(t)+\sigma_j(t))^2}} \geq d_i \right\} \tag{10}$$

and the uncertainty input

$$u_i(t+1) = b \left| \bar{x}_i(t) - \frac{1}{|N_i(t)|} \sum_{j \in N_i(t)} (\bar{x}_j(t)) \right| \tag{11}$$

for *local reference scheme*, or

$$u_i(t+1) = b|\bar{x}_i(t) - g_i(t)| \tag{12}$$

for *external reference scheme* with initial condition $\bar{x}_i(0) = x_{i0}$ (initial opinion of agent $i$) and $\sigma_i(0) = \sigma_{i0}$ (uncertainty about the initial opinion), where $0 \leq d_i \leq 1$, $b > 0$ are constants. ∎

We now define the basic leader-follower group which is the basic building block of the top-down hierarchical fuzzy opinion networks of this paper.

**Definition 3:** A *basic leader-follower group* (BLFG), illustrated in Fig. 1, consists of a leader node $X_a(t)$ with membership function $\mu_{X_a(t)}(x) = e^{-\frac{|x-\bar{x}_a(t)|^2}{(\sigma_a(t))^2}}$ and $n$ follower nodes $X_i(t)$ ($i = 1, 2, \ldots, n$) with membership functions $\mu_{X_i(t)}(x) = e^{-\frac{|x-\bar{x}_i(t)|^2}{(\sigma_i(t))^2}}$, where the leader node passes his opinion $\bar{x}_a(t)$ to each of the $n$ follower nodes and the $n$ follower nodes are connected among themselves in the bounded confidence fashion. Specifically, the leader's opinion $\bar{x}_a(t)$ and its uncertainty $\sigma_a(t)$ are not influenced by the $n$ followers, and the opinions $\bar{x}_i(t)$ and their uncertainties $\sigma_i(t)$ of the $n$ followers are evolving according to the following dynamic equations:

$$\bar{x}_i(t+1) = \frac{1}{|N_i(t)|+1} \left( \sum_{j \in N_i(t)} (\bar{x}_j(t)) + \bar{x}_a(t) \right) \tag{13}$$

$$\sigma_i(t+1) = \frac{1}{|N_i(t)|} \sum_{j \in N_i(t)} (\sigma_j(t)) + u_i(t+1) \tag{14}$$

where $N_i(t)$ is given in (10) with $0 \leq d_i < 1$, and the uncertainty input $u_i(t+1)$ is chosen either with the *local reference scheme* (11), or with the *leader reference scheme*:

$$u_i(t+1) = b|\bar{x}_i(t) - \bar{x}_a(t)| \tag{15}$$

∎

We see from (13) that the opinion $\bar{x}_i$ of follower $i$ is updated as the average of the neighbor's opinions $\sum_{j \in N_i(t)} (\bar{x}_j(t))$ plus the leader's opinion $\bar{x}_a(t)$. For the uncertainty $\sigma_i$ of follower $i$, we see from (14) that it is updated as the average of the neighbor's uncertainties $\frac{1}{|N_i(t)|} \sum_{j \in N_i(t)} (\sigma_j(t))$ plus the uncertainty input $u_i(t+1)$ which takes either the local reference scheme (11) or the leader reference scheme (15). In the local reference scheme, agent $i$ views the average of his neighbor's opinions as the reference, so the closer his opinion is to this average, the less uncertainty he has. In the leader reference scheme, however, agent $i$ views the leader's opinion as the reference, so the closer his opinion $\bar{x}_i(t)$ is to the leader's opinion $\bar{x}_a(t)$, the less uncertainty he has.

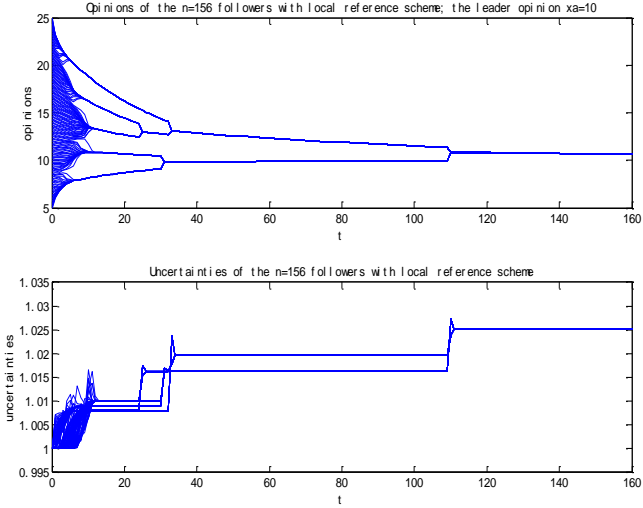

Fig. 2: A simulation run of the basic leader-follower group with local reference scheme, where the top and bottom sub-figures plot the opinions $\bar{x}_i(t)$ and the uncertainties $\sigma_i(t)$ of the $n$=156 followers, respectively; the leader's opinion $\bar{x}_a(t) = 10$.

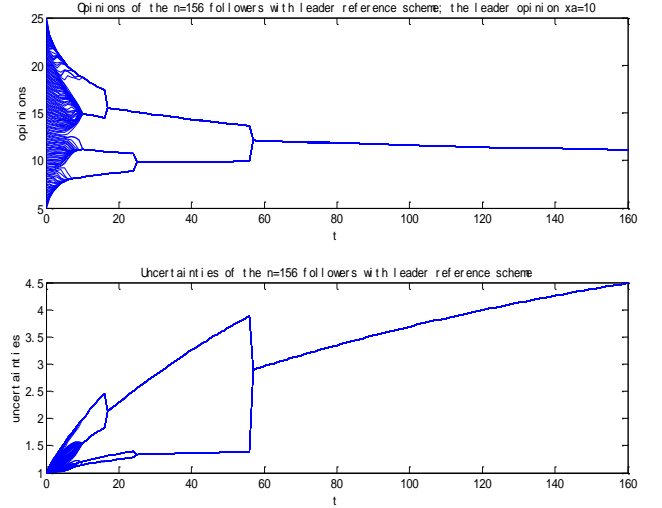

Fig. 3: A simulation run of the basic leader-follower group with leader reference scheme, where the top and bottom sub-figures plot the opinions $\bar{x}_i(t)$ and the uncertainties $\sigma_i(t)$ of the $n$=156 followers, respectively; the leader's opinion $\bar{x}_a(t) = 10$.

To get a feel of the dynamics of the opinions and their uncertainties of the agents in the basic leader-follower group, let's see an example.

**Example 1**: Consider the basic leader-follower group of Fig. 1 with $n = 156$ followers. With $d_i = 0.6$, $b = 0.01$, the leader's opinion $\bar{x}_a(t) = 10$ and the initial $x_{i0}$ ($i = 1, ..., n$) uniformly distributed over the interval [5,25] ($x_{i0} = 5 + 20(\frac{i-1}{n-1})$, $i = 1, ..., n$) and their uncertainties $\sigma_{i0} = 1$ for all $i = 1, ..., n$, Fig. 2 and Fig. 3 show the simulation runs of the dynamic model with local reference scheme (11) and leader reference scheme (15), respectively, where the top sub-figures of Figs. 2 and 3 plot the opinions $\bar{x}_i(t)$ of the $n = 156$ followers and the bottom sub-figures plot the uncertainties $\sigma_i(t)$. ∎

We see from Figs. 2 and 3 that for both the local and leader reference schemes, the opinions $\bar{x}_i(t)$ of all the followers converge to the leader's opinion $\bar{x}_a(t) = 10$, but the speed of convergence is slow. In the following theorem, we prove that convergence to the leader's opinion is indeed guaranteed, but the speed of convergence is greatly influenced by the number of followers in the group.

**Theorem 1:** Consider the basic leader-follower dynamics of (13), (14) and (10) with local reference scheme (11) or leader reference scheme (15), and suppose the leader's opinion $\bar{x}_a(t) = \bar{x}_a$ is a constant. Starting from arbitrary initial opinions $\bar{x}_i(0) = x_{i0} \in R$ and uncertainties $\sigma_i(0) = \sigma_{i0} \in R_+$, we have:

(a) the $n$ followers converge to a consensus in finite time, i.e., there exists $t_N$ such that $\bar{x}_i(t) = \bar{x}(t)$ and $\sigma_i(t) = \sigma(t)$ for all $i = 1, ..., n$ and all $t > t_N$;

(b) the opinion consensus $\bar{x}(t)$ converges to the leader's opinion $\bar{x}_a$ according to

$$\bar{x}(t) = \bar{x}_a + \left(\frac{n}{n+1}\right)^{t-t_N-1} (\bar{x}(t_N + 1) - \bar{x}_a) \quad (16)$$

where $t > t_N + 1$;

(c) for local reference scheme (11), the uncertainty consensus $\sigma(t) = \sigma(t_N + 1)$ (a constant) for all $t > t_N$;

(d) for leader reference scheme (15), the uncertainty consensus $\sigma(t)$ is changing according to

$$\sigma(t) = \sigma(t_N + 1)$$
$$+ b|\bar{x}(t_N + 1) - \bar{x}_a| \sum_{k=t_N+1}^{t} \left(\frac{n}{n+1}\right)^{k-t_N-1} \quad (17)$$

for $t > t_N + 1$, from which we get $\lim_{t\to\infty} \sigma(t) = \sigma(t_N + 1) + b|\bar{x}(t_N + 1) - \bar{x}_a|(n+1)$. ∎

The proof of Theorem 1 is given in the Appendix.

From (16) in Theorem 1 we see that the opinion consensus $\bar{x}(t)$ converges to the leader's opinion $\bar{x}_a$ with the factor $\frac{n}{n+1}$, i.e., the error $\bar{x}(t) - \bar{x}_a$ is reduced by $\frac{n}{n+1}$ each time step, so in $k$ time steps the error $\bar{x}(t) - \bar{x}_a$ is reduced by $\left(\frac{n}{n+1}\right)^k = \varepsilon$ which gives

$$k = \frac{\log \varepsilon}{\log n - \log(n+1)} \quad (18)$$

With $\varepsilon = 0.01$ (reduce to error to 1%), Fig. 4 plots the $k$ as function of $n$, from which we see that the steps needed to reduce the error increases about linearly with the number of followers in the group, meaning that larger groups are more difficult to converge to the leader's opinion than smaller groups.

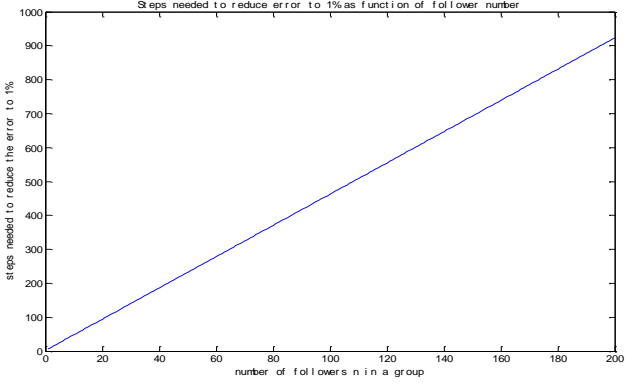

Fig. 4: Plot of (18), the steps $k$ needed to reduce the error $\bar{x}(t) - \bar{x}_a$ to 1% as function of number of followers $n$ in group.

The conclusion from (18) and Fig. 4 is that to speed up the convergence of the followers' opinions to the leader's opinion, reducing the size of the group is crucial. Organizing the followers hierarchically in smaller groups, as we will do next through the top-down hierarchical fuzzy opinion networks, is an efficient way to speed up the convergence.

We now introduce the top-down hierarchical fuzzy opinion networks.

**Definition 4:** A *top-down hierarchical fuzzy opinion network* (TD-HFON), illustrated in Fig. 5, is constructed from a number of basic leader-follower groups of Fig. 1 in a multi-layer structure, where an agent $X_{ij}^l$ in Level $l$ is a follower to an agent in Level $l+1$ and is a leader to some agents in Level $l$-1. In the notation $X_{ij}^l$, $l$ is the level index ($l = 1,2, \ldots, L$), $i$ is the group index ($i = 1,2, \ldots, n^l$), $j$ is the index in the group ($j = 1,2, \ldots, n_i^l$), and $X_{ij}^l$ is a Gaussian fuzzy set with center $\bar{x}_{ij}^l(t)$ and standard deviation $\sigma_{ij}^l(t)$. ∎

To see how fast the hierarchical structure can speed up the convergence to the leader's opinion, we reorganize the $n$=156 followers in Example 1 into a 3-level and a 4-level TD-HFONs in the following example.

**Example 2:** Consider the 3-level and 4-level TD-HFONs in Fig. 6. In the 3-level TD-HFON (left in Fig. 6), Level-1 consists of 12 groups with 12 agents in each group, Level-2 consists of a single group of 12 agents with each agent being the leader of one the 12 groups in Level-1, and Level-3 is the top leader who is the leader of the 12-agent group in Level-2. With $d_i = 0.6$, $b = 0.01$, the top leader's opinion $\bar{x}_a(t) = 10$ and the initial opinions of the 12 agents in a group $\bar{x}_{ij}^l(0)$ ($l = 1$ or $2, i = 1, \ldots, 12$ for $l = 1$, and $j = 1, \ldots, n$) uniformly distributed over the interval [5,25] ($\bar{x}_{ij}^l(0) = 5 + 20(\frac{j-1}{12-1}), j = 1, \ldots, 12$) and all their uncertainties $\sigma_{ij}^l(0) = 1$, Figs. 7 and 8 show the simulation runs of the dynamic model with local reference scheme (11) and leader reference scheme (15), respectively, where the top sub-figures of Figs. 7 and 8 plot the opinions $\bar{x}_{ij}^l(t)$ of the $n = 156$ agents in Levels 1 and 2 and the bottom sub-figures plot the corresponding uncertainties $\sigma_{ij}^l(t)$.

Similarly, in the 4-level TD-HFON (right in Fig. 6), Level-1 consists of 25 groups with 5 agents in each group, Level-2 consists of 5 groups with 5 agents in each group and these 25 agents are the leaders of the 25 groups in Level-1, Level-3 consists of a single group of 5 agents who are the leaders of the 5 groups in Level-2, and Level-4 is the top leader who is the leader of the 5-agent group in Level-3. With $d_i = 0.6$, $b = 0.01$, the top leader's opinion $\bar{x}_a(t) = 10$ and the initial opinions of the 5 agents in a group $\bar{x}_{ij}^l(0)$ ($l = 1, 2$ or $3, i = 1, \ldots, 5$ for $l = 2$ and $i = 1, \ldots, 25$ for $l = 1$, and $j = 1, \ldots, 5$) uniformly distributed over the interval [5,25] ($\bar{x}_{ij}^l(0) = 5 + 20(\frac{j-1}{5-1}), j = 1, \ldots, 5$) and all their uncertainties $\sigma_{ij}^l(0) = 1$, Figs. 9 and 10 show the simulation runs of the dynamic model with local reference scheme (11) and leader reference scheme (15), respectively, where the top sub-figures of Figs. 9 and 10 plot the opinions $\bar{x}_{ij}^l(t)$ of the $n = 155$ agents in Levels 1, 2 and 3 and the bottom sub-figures plot the corresponding uncertainties $\sigma_{ij}^l(t)$. ∎

Comparing Fig. 2 and 3 with Figs. 7-10, we have the following observations:

(a) The opinions $\bar{x}_{ij}^l(t)$ of all the agents converge to the leader's opinion no matter the agents are organized hierarchically in small groups or in one large group.

(b) The speed of convergence to the leader's opinion is greatly improved when the agents are organized hierarchically in small groups; the more the levels or the smaller the groups, the faster the convergence will be (comparing the top sub-figures of Figs. 2, 7 and 9 for the local reference scheme, and the top sub-figures of Figs. 3, 8 and 10 for the leader reference scheme).

(c) Although the opinions $\bar{x}_{ij}^l(t)$ of all the agents converge to the leader's opinion, their uncertainties $\sigma_{ij}^l(t)$ in general converge to different values for agents in different groups, reflecting the different processes that the agents in different groups were experiencing during the convergence to the leader's opinion.

Indeed, we will prove in the following theorem that the observations above are true in general.

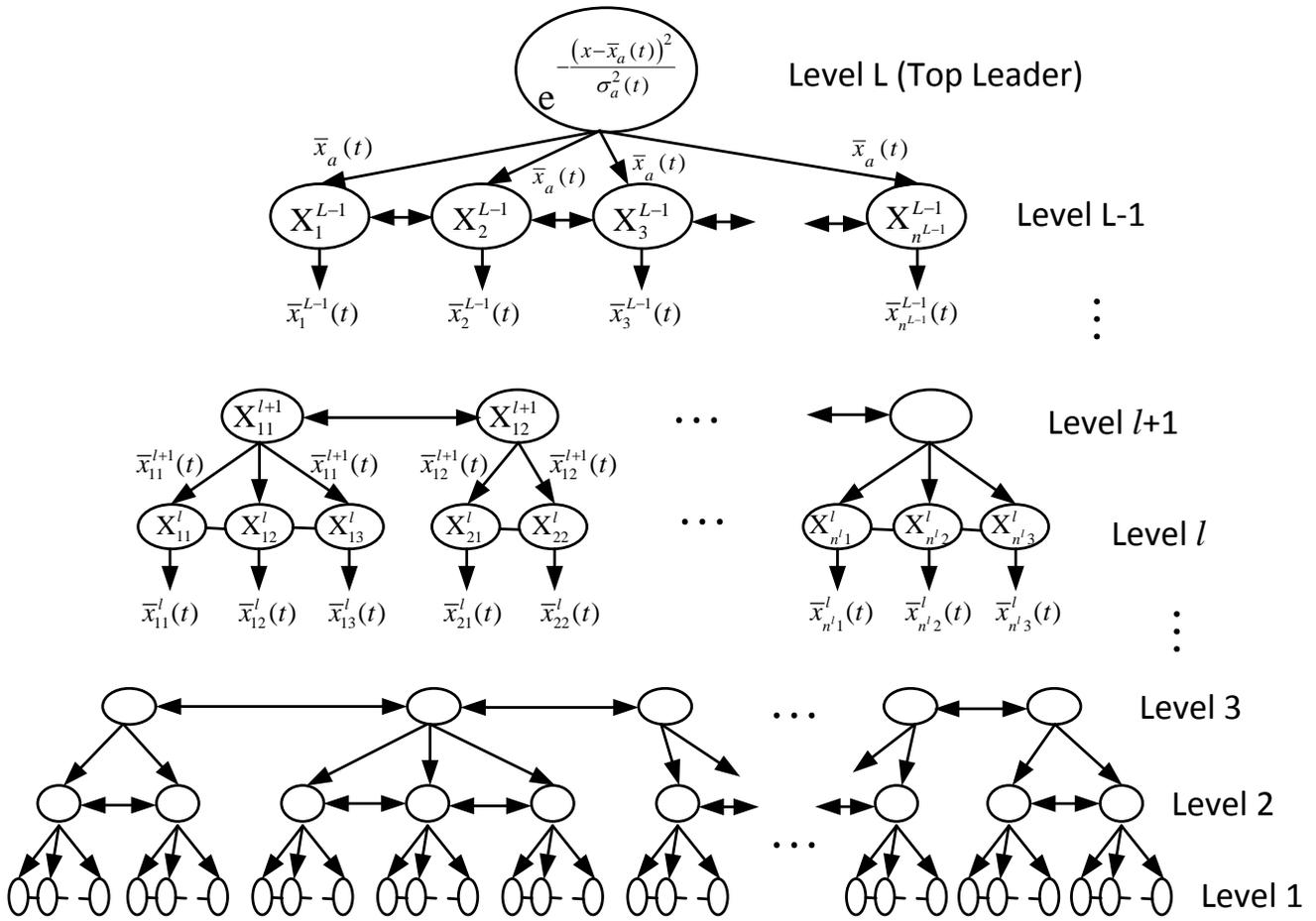

Fig. 5: The top-down hierarchical fuzzy opinion networks.

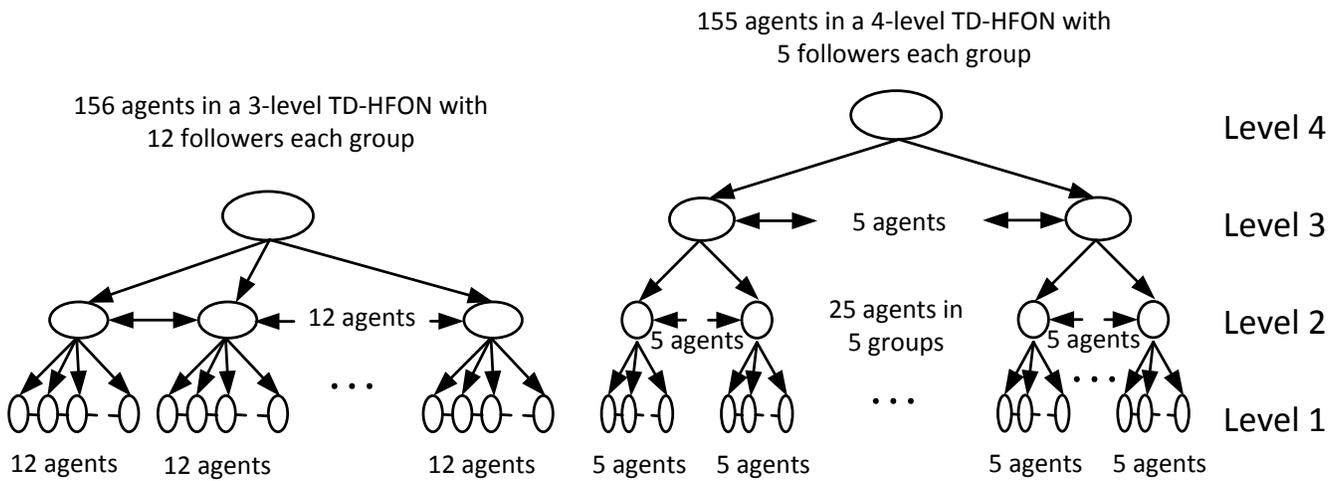

Fig. 6: Reorganizing the agents in Example 1 into a 3-level TD-HFON with 12 followers in each group (left) and a 4-level TD-HFON with 5 followers in each group (right).

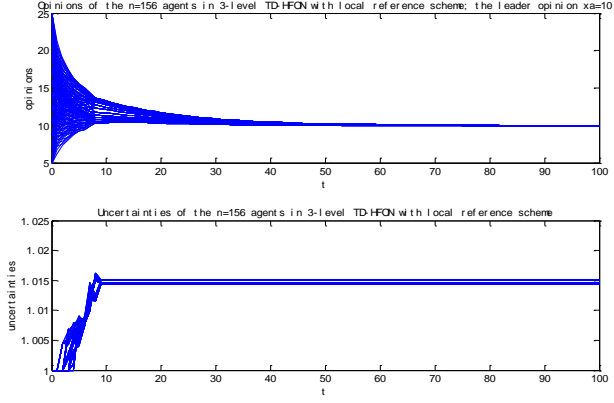

Fig. 7: The opinions (top) and their uncertainties (bottom) of the n=156 agents in the 3-level TD-HFON of Fig. 6 with local reference scheme.

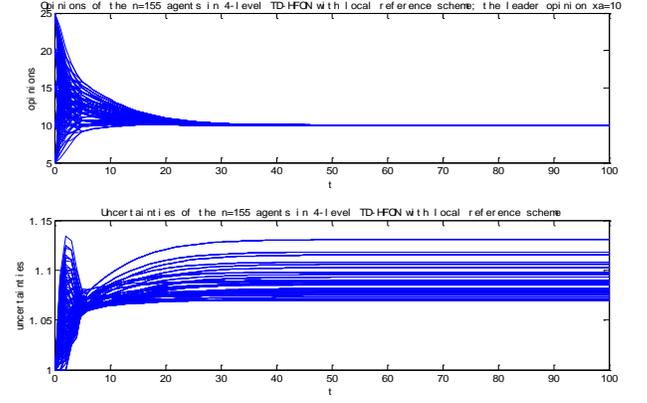

Fig. 9: The opinions (top) and their uncertainties (bottom) of the n=155 agents in the 4-level TD-HFON of Fig. 6 with local reference scheme.

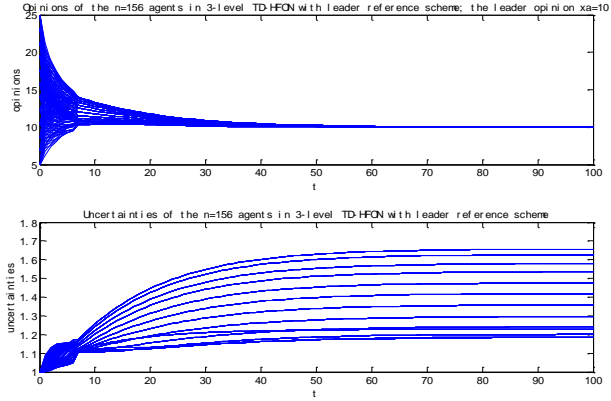

Fig. 8: The opinions (top) and their uncertainties (bottom) of the n=156 agents in the 3-level TD-HFON of Fig. 6 with leader reference scheme.

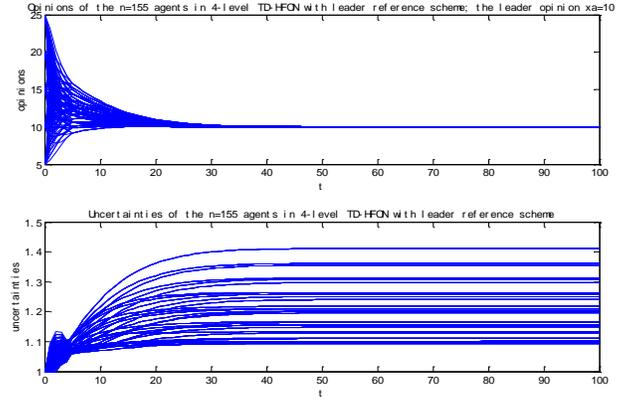

Fig. 10: The opinions (top) and their uncertainties (bottom) of the n=155 agents in the 4-level TD-HFON of Fig. 6 with leader reference scheme.

**Theorem 2:** Consider the general TD-HFON in Fig. 5 with dynamics of all the groups following (13), (14) and (10) with local reference scheme (11) or leader reference scheme (15), and suppose the top leader's opinion $\bar{x}_a(t) = \bar{x}_a$ is a constant. Starting from arbitrary initial opinions $\bar{x}_{ij}^l(0) \in R$ and uncertainties $\sigma_{ij}^l(0) \in R_+$, we have:

(a) the opinions $\bar{x}_{ij}^l(t)$ of all the agents ($l = 1, 2, \ldots, L-1$; $i = 1, 2, \ldots, n^l$; $j = 1, 2, \ldots, n_i^l$) converge to the leader's opinion $\bar{x}_a$;

(b) the uncertainties $\sigma_{ij}^l(t)$ of the followers in the same leader-follower group converge to a constant, but different groups in general converge to different values. ∎

The proof of Theorem 2 is given in the Appendix.

We now move to the next section to study the bottom-up hierarchical fuzzy opinion networks.

### III. BOTTOM-UP HIERARCHICAL FUZZY OPINION NETWORKS

As we discussed in the Introduction that although social hierarchy is prevalent throughout culture and time, hierarchy is against some of the best values of humanity – hierarchy is undemocratic, unequal, unfair, and unjust. So, if we have to choose hierarchy to govern a large population such as a nation, we should have some counter measures to prevent those in the higher levels to abuse their power. Election by the general public is the way of choice of most countries in the world to select their top leaders. In the election scenario, the opinions of the large population are initially very diversified and many small leaders are emerging to represent different interest groups, then these small leaders have to compromise with each other to select the middle-level leaders, this process continues level-by-level in a bottom-up fashion until some consensuses are reached. We now propose the bottom-up hierarchical fuzzy opinion networks to model such processes.

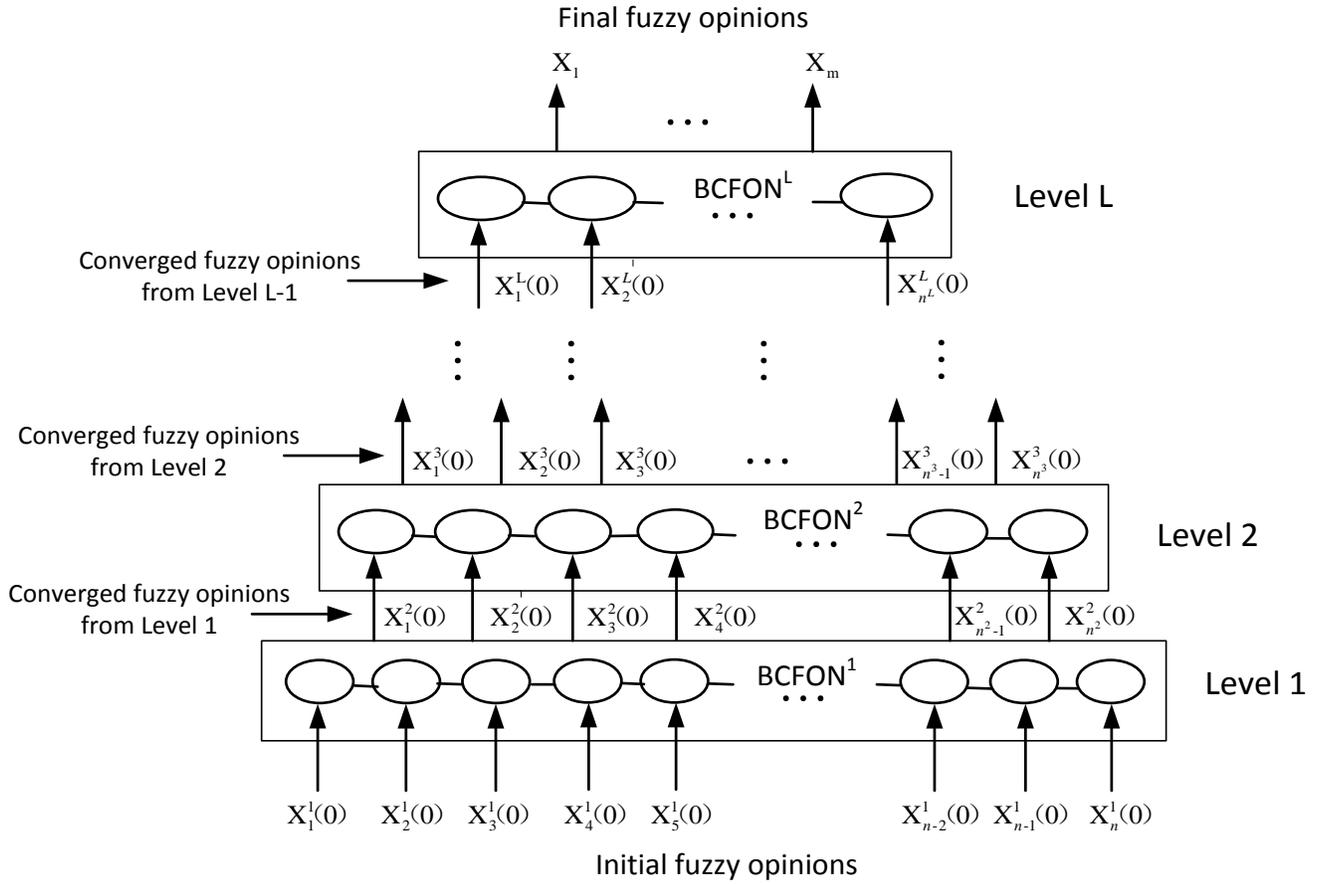

Fig. 11: The bottom-up hierarchical fuzzy opinion networks.

**Definition 5:** A *bottom-up hierarchical fuzzy opinion network* (BU-HFON), illustrated in Fig. 11, is the layered connection of a number of bounded confidence fuzzy opinion networks (BCFON) with local reference scheme (Definition 2), where the converged opinions of a lower level BCFON are passed to the upper level BCFON as the initial opinions. ∎

We now simulate the BU-HFON to see how the initial fuzzy opinions are agglomerated layer-by-layer in some typical situations.

**Example 3:** Consider a 5-level BU-HFON of Fig. 11 ($L$=5) with $n = 200$ agents in Level 1 whose initial opinions $\bar{x}_i^1(0)$ and initial uncertainties $\sigma_i^1(0)$ ($i$=1,2, …, $n$) are randomly distributed over the intervals [5,25] and (0,1), respectively. The five BCFONs in the five levels are evolving according to the dynamic equations (7)-(11), where $d_i = 0.95$ for Level 1 BCFON, $d_i = 0.7$ for Level 2 BCFON, $d_i = 0.45$ for Level 3 BCFON, $d_i = 0.2$ for Level 4 BCFON and $d_i = 0.05$ for Level 5 BCFON. The meaning of these $d_i$'s are explained as follows.

For the Level 1 agents (the general public), we choose a large $d_i$ (=0.95) because the general public has no obligation to reach some consensuses so that they can show little sign of compromise (a large $d_i$ means talking only to those whose opinions are very close to each other). For the Level 2 agents (the local representatives of the general public), they have to show some sign of compromise in order for the process to continue, so we choose a little smaller $d_i$ (=0.7) to model the situation. Then, the Level 3 agents must be even more compromising in order to reach some rough consensuses, so we choose a even smaller $d_i$ (=0.45) for these middle level agents. This process continues with smaller and smaller $d_i$'s for the upper level agents ($d_i = 0.2$ for Level 4 and $d_i = 0.05$ for Level 5) because the higher the level they are in, the more pressure they have to reach the final consensus (this is why many elected agents fall to realize their election promises when they are in the office, because they have to consider many different concerns when they are in the higher levels).

With $b$=0.5 for all the BCFONs and each BCFON evolving 40 time steps, i.e., the Level 1 BCFON is operating from $t$=0 to $t$=40, then followed by the Level 2 BCFON which is operating from $t$=41 to $t$=80 with the converged fuzzy opinions of the Level 1 BCFON as the initial values, this process continues with the Level 3 BCFON operating from t=81 to t=120, the Level 4 BCFON operating from t=121 to t=160 and the Level 5 BCFON operating from t=160, Fig. 12 shows a simulation run in a typical situation, where the top and bottom sub-figures in

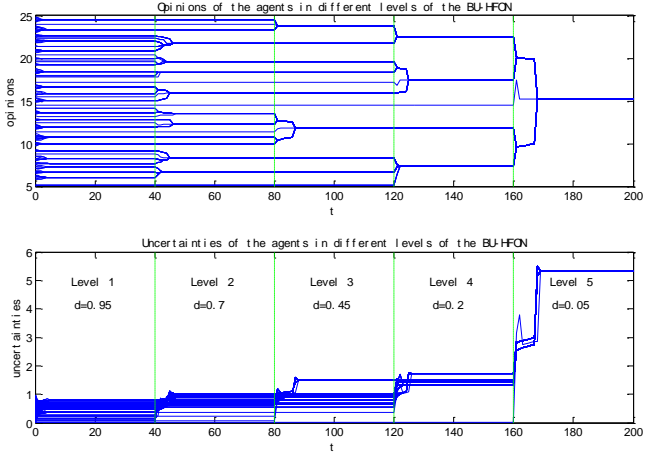

Fig. 12: The opinions (top) of the agents in different levels and their uncertainties (bottom).

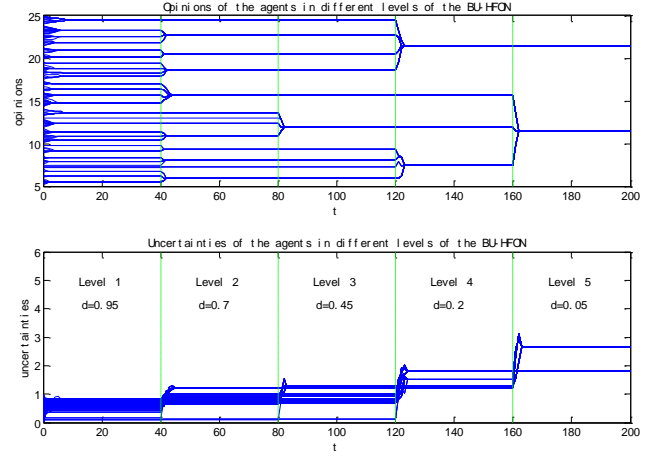

Fig. 14: The opinions (top) of the agents in different levels and their uncertainties (bottom).

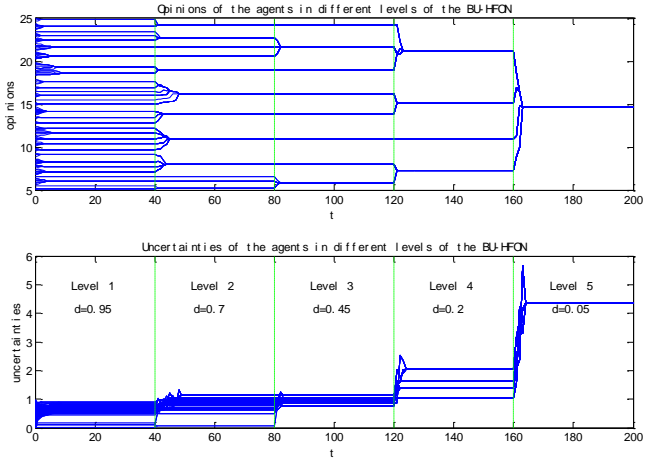

Fig. 13: The opinions (top) of the agents in different levels and their uncertainties (bottom).

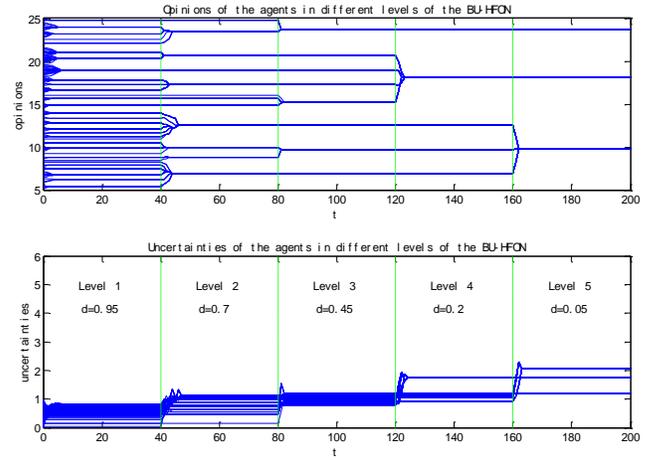

Fig. 15: The opinions (top) of the agents in different levels and their uncertainties (bottom).

Fig. 12 show the opinions ($\bar{x}_i(t)$ of (7)) of the agents and their uncertainties ($\sigma_i(t)$ of (8)), respectively. We see from top sub-figure of Fig. 12 that the Level 1 general public ($n = 200$) converge to a large number of opinions due to the large $d_i$ (=0.95), then with a smaller $d_i$ (=0.7) the Level 2 agents converge to about 17 opinions, which are further combined by the Level 3 agents (with $d_i = 0.45$) into 11 opinions, and continuing with $d_i = 0.2$ the Level 4 agents reach 5 opinions, finally, the top level agents have to adopt a very small $d_i = 0.05$ to reach a single consensus. The bottom sub-figure of Fig. 12 shows that the uncertainties are getting larger and larger for the higher level agents, reflecting the fact that the higher level agents must demonstrate more compromises which result in more uncertainties about their decisions.

Figs. 13-15 show the simulation runs in other typical situations, where a consensus is reached in Fig. 13, but in the situations of Figs. 14 and 15, a consensus cannot be reached after five rounds of negotiations. Comparing the bottom sub-figures of Figs. 12-15 we see that the uncertainties of the Level 5 agents are high if they converge to a single consensus (Figs. 12 and 13), but if they converge to two consensuses (Fig. 14), their uncertainties are much lower, and if they are allowed to keep three different opinions (Fig. 15), their uncertainties are even lower. This demonstrates that the uncertainty $\sigma_i(t)$ in our HFON model provides a good measure for the psychological pressures of the agents in different levels. ∎

## IV. CONCLUDING REMARKS

The top-down and bottom-up hierarchical fuzzy opinion networks (HFON) developed in this paper provide a mathematical framework to model the dynamical propagation and formation of opinions and their uncertainties through the hierarchical structures. For the top-down HFON, we prove that the opinions of all followers throughout the hierarchy converge to the top leader's opinion, but the uncertainties of the followers in different groups are different, which means that although all the followers have to follow the top leader's opinion, their psychological acceptance (the uncertainty) for the top leader's opinion is different. We show that the iterations needed to reduce the tracking error between the followers and leader's opinions by a certain percentage is proportional to the number of followers in the group, this means that organizing the followers hierarchically can greatly improve the efficiency; for example, if we organize 155 followers in a 4-level top-down HFON with five followers in each leader-follower group, then the speed of convergence to the top leader's opinion is approximately $\frac{155}{5*3} \approx 10$ times faster than organizing the 155 followers in a single flat group. For the bottom-up HFON, we show that the psychological pressure (the uncertainty) of the agents in the higher levels is greater than those in the lower levels because the higher level agents have to make more compromises (tougher decisions), also we show that the uncertainties are lower if the higher level agents are allowed to keep different opinions.

In the future research, we will apply the HFON models to some real organizations and real election scenarios.

## APPENDIX

**Proof of Theorem 1:** (a) Let $X(t) = (\bar{x}_1(t), \dots, \bar{x}_n(t), \bar{x}_a(t))^T$ and $W(t) = [w_{ij}(t)]_{(n+1) \times (n+1)}$ with

$$w_{ij}(t) = \begin{cases} \frac{1}{|N_i(t)|+1}, & j \in N_i(t) \text{ or } j = n+1 \\ 0, & j \notin N_i(t) \end{cases} \quad (A1)$$

for $i = 1, \dots, n$, $j = 1, \dots, n+1$,

$$w_{(n+1)j}(t) = 0 \quad (A2)$$

for $j = 1, \dots, n+1$ and

$$w_{(n+1)(n+1)}(t) = 1 \quad (A3)$$

Then, with the leader's opinion $\bar{x}_a(t) = \bar{x}_a$ being a constant, the opinion dynamic equation (13) can be rewritten in the matrix form:

$$X(t+1) = W(t) X(t) \quad (A4)$$

We need the follow Lemma from [27] to continue our proof.

**Lemma**: If the row-stochastic matrix $W(t)$ in (A4) satisfies the following three conditions:

i) the diagonal of $W(t)$ is positive, i.e., $w_{ii}(t) > 0$ for $i = 1, \dots, n+1$,

ii) there is $\delta > 0$ such that the lowest positive entry of $W(t)$ is greater than $\delta$, and

iii) any two nonempty saturated sets for $W(t)$ have a nonempty intersection, where $I \subseteq \{1, \dots, n+1\}$ is saturated for $W(t)$ if $w_{ij}(t) > 0$ and $i \in I$ implies $j \in I$, then a consensus is reached for $\bar{x}_1(t), \dots, \bar{x}_n(t)$ in finite time.

We now show that the $W(t)$ of (A1)-(A3) satisfies the three conditions in the Lemma. Since $i \in N_i(t)$ according to the definition of $N_i(t)$ in (10), we have $w_{ii}(t) = \frac{1}{|N_i(t)|+1} > 0$ for $i = 1, \dots, n$; with $w_{(n+1)(n+1)}(t) = 1$, condition i) in the Lemma is satisfied. Since $|N_i(t)| \leq n$, it follows that any positive $w_{ij}(t) = \frac{1}{|N_i(t)|+1} > \frac{1}{n+2} \equiv \delta$, hence condition ii) of the Lemma is satisfied. To check condition iii), notice from (A1) that $w_{i(n+1)}(t) = \frac{1}{|N_i(t)|+1} > 0$ for $i = 1, \dots, n$, which implies that any two nonempty saturated sets $I_1, I_2 \subseteq \{1, \dots, n+1\}$ for $W(t)$ must contain the element $n+1$, hence condition iii) of the Lemma is satisfied. Consequently, according to the Lemma, the $n$ followers $\bar{x}_1(t), \dots, \bar{x}_n(t)$ converge to a consensus in finite time, i.e., there exists $t_N$ such that $\bar{x}_i(t) = \bar{x}(t)$ for all $i = 1, \dots, n$ and all $t > t_N$.

To prove $\sigma_i(t) = \sigma(t)$ for $i = 1, \dots, n$ and $t > t_N$, notice that for $t > t_N$, $u_i(t+1) = 0$ for the local reference scheme (11), and $u_i(t+1) = b|\bar{x}(t) - \bar{x}_a|$ for the leader reference scheme (15). Substituting these $u_i(t+1)$ into the dynamic equation (14) of $\sigma_i(t)$, we have for $t > t_N$ that

$$\sigma_i(t+1) = \frac{1}{n} \sum_{j=1}^{n} \sigma_j(t) \quad (A5)$$

for the local reference scheme (11), and

$$\sigma_i(t+1) = \frac{1}{n} \sum_{j=1}^{n} \sigma_j(t) + b|\bar{x}(t) - \bar{x}_a| \quad (A6)$$

for the leader reference scheme (15). Since the right hand sides of both (A5) and (A6) are independent of $i$, we have in both cases that $\sigma_i(t) = \sigma(t)$. This completes the proof of (a) of Theorem 1.

(b) Since $\bar{x}_i(t) = \bar{x}(t)$ for all $i = 1, \dots, n$ when $t > t_N$, we have from (13) that

$$\bar{x}(t+1) = \frac{n}{n+1} \bar{x}(t) + \frac{1}{n+1} \bar{x}_a \quad (A7)$$

or

$$\bar{x}(t+1) - \bar{x}_a = \frac{n}{n+1} (\bar{x}(t) - \bar{x}_a) \quad (A8)$$

for $t > t_N$, and (16) follows from (A8).

(c) The conclusion follows from (A5).

(d) Substituting $\sigma_i(t) = \sigma(t)$ into (A6), we have

$$\sigma(t+1) = \sigma(t) + b|\bar{x}(t) - \bar{x}_a| \quad (A9)$$

for $t > t_N$, and (17) follows from (A9) and (16). ∎

**Proof of Theorem 2:** (a) Consider an arbitrary leader-follower group in the HFON and let $\bar{x}_i^l(t)$ be the group leader's opinion and $\bar{x}_{ij}^{l-1}(t)$ $(j = 1,2, \ldots, n_i^l)$ be the followers' opinions. Since all the $n_i^l$ followers are connected to each other through the group leader (the group leader is a common element in any saturated set), we have from the Lemma that the followers converge to a consensus in finite steps, i.e., there exists $t_N$ such that $\bar{x}_{ij}^{l-1}(t) = \bar{x}_i^{l-1}(t)$ when $t > t_N$, so from (13) we have

$$\bar{x}_i^{l-1}(t+1) = \frac{n_i^l}{n_i^l + 1} \bar{x}_i^{l-1}(t) + \frac{1}{n_i^l + 1} \bar{x}_i^l(t) \quad \text{(A10)}$$

or

$$\bar{x}_i^{l-1}(t+1) - \bar{x}_a = \frac{n_i^l}{n_i^l + 1} \big(\bar{x}_i^{l-1}(t) - \bar{x}_a\big) + \frac{1}{n_i^l + 1} \big(\bar{x}_i^l(t) - \bar{x}_a\big) \quad \text{(A11)}$$

for $t > t_N$. If $\bar{x}_i^l(t)$ converges to $\bar{x}_a$, then since $\frac{n_i^l}{n_i^l+1} < 1$ we have from (A11) that $\bar{x}_i^{l-1}(t)$ converges to $\bar{x}_a$, i.e., if the group leader's opinion converges to the top leader's opinion $\bar{x}_a$, then consensus of the followers in the group also converges to $\bar{x}_a$. Since the top leader and the agents in Level L-1 form a basic leader-follower group, we have from Theorem 1 that the opinions of the agents in Level L-1 converge to the top leader's opinion $\bar{x}_a$. Hence, by induction, we have that the opinions of all the agents converge to the top leader's opinion $\bar{x}_a$.

(b) For the agents in the same group (say group $i$ in Level $l$), we can use the same method as in the proof of Theorem 1 to show that their uncertainties $\sigma_{ij}^l(t)$ $(j = 1,2, \ldots, n_i^l)$ reach a consensus $\sigma_i^l(t)$ in finite time which converges to the constant $\sigma_i^l(t_N + 1)$ for the local reference scheme or to the constant $\sigma_i^l(t_N + 1) + b|\bar{x}_i^l(t_N + 1) - \bar{x}_a|(n_i^l + 1)$ for the leader reference scheme. Since these converged values are group dependent, they are in general different for different groups. ∎